# Two-pulse stimulated echo in magnets

M.D. Zviadadze[a], G.I. Mamniashvili[a], T.O. Gegechkori[a], A.M. Akhalkatsi[b], T.A. Gavasheli[b]

[a] Javakhishvili Tbilisi State University, Andronikashvili Institute of Physics,
6 Tamarashvili st. Tbilisi 0177, Georgia
[b] I.Javakhishvili Tbilisi State University, 3 Chavchavadze av. 0128. Tbilisi, Georgia
m.zviadadze@mail.ru

## Summary

The results of experimental study of two-pulse stimulated echo in ferromagnets of two types are presented. Ferromagnet $Co$ and half-metal $Co_2MnSi$, in which a single-pulse echo formed by the distortion mechanism of the fronts of exciting pulse is also observed, are classified among the first type. Lithium ferrite and intermetal compound $MnSb$ characterized by the absence of single-pulse echo in them – belong to the second type.

For signals of two-pulse stimulated echo in the materials of the first type a short time and a long time of relaxations are observed. The short time is about the order of value shorter less than the spin-spin relaxation time. The long time is close to the transverse relaxation time of single-pulse echo formed by the distortion mechanism. The mechanisms that provide the possible interpretations of the peculiarities of the processes of nuclear magnetic relaxation are discussed.

# INTRODUCTION

One of the main ways to measure of the longitudinal (spin-lattice) relaxation $T_1$ and investigation of spin-diffusion processes is the method of stimulated spin echo [1]. In the traditional Hahn mechanism of spin-echo formation for the observation of stimulated echo it is necessary to have at least three resonant RF pulses [1]. The possibility to observe the stimulated echo is related as it is known with fact that the $z$-component of magnetization of $j$-th isochromate after the termination of the second RF pulse depends on the phase of its transverse component at the start of the second pulse $\varphi_j = \Delta\omega_j \cdot \tau_{12}$, where $\Delta\omega_j$ is the resonant frequency of $j$-th isochromate in the system of coordinates rotating with the RF pulse filling frequency $\omega_{RF}$, $\tau_{12}$ - is the interval between the first and the second RF pulses.

In work [2] was developed the SPE theory for Hahn spin systems was developed, i.e. the pulse response of spin-system occurring at the action of one wide RF pulse at time approximately equal to the duration of the exciting RF pulse.

In following the theory was developed in a number of works, among which let us note [3-7] ones. It was shown that using the SPE method one could obtain practically the same information on spin-system as by two-pulse echo (TPE) Hahn method. But it was also noted [2] that the transverse relaxation time $T_2^S$, defined using the TPE method due to the difference of effective magnetic fields in the rotating coordinate system (RCS) in the dephasing and rephasing processes of nuclear isochromates, as result of which angular rates of $j$-th isochromate precession during the RF action and after its termination are different and equal to, correspondingly, $\Delta\omega_j' = (\Delta\omega_j^2 + \omega_1^2)^{1/2}$ and $\Delta\omega_j = \omega_j - \omega_{RF}$, $\omega_j$ is the NMR frequency of $j$-th isochromate, $\omega_1 = \gamma_n \eta H_1$ is the pulse amplitude in the frequency units, $\eta$ is the enhancement factor and $\gamma_n$ is the gyromagnetic ratio. Due to the $\Delta\omega_j \neq \Delta\omega_j'$ inequality the SPE signal could be observed only at fulfillment of specific conditions, pointed in [2]. Besides it, the transverse relaxation time $T_2^S$ could turn out to be shorter than $T_2$ measured by the TPE method.

The formation mechanism of SPE makes it possible also the observation of TPSE in case of application of a short additional (reading) RF pulse [2,8]. The point is in the fact that during the RF pulse action the inclination of isochromate precession axis (that causes the trajectory in the $xy$ plane to be ellipsoidal and produces the SPE signal) makes also the $z$-component of the isochromate dependent on the phase of its transverse component at the install of termination of the first pulse $\varphi_j' = \Delta\omega_j' \cdot \tau_1$. Thus, the difference of inclinations of effective magnetic fields in the PCS during the action of after the termination of RF pulse causes the formation of the SPE and the possibility to observe the TPSE using only two, wide and short, exciting RF pulses.

Following to [2, 6], it is easy to find out the expression for the TPSE intensity.

Assume system of nuclear spins in an equilibrium state is excited by the RF pulse of duration $\tau_1$, amplitude $H_1$ and frequency $\omega_{RF}$. The motion of $j$-th isochromate magnetization vector $\vec{m}_j$ in RCS constitutes then precession around an effective field

$$\vec{H}_e = \gamma_n^{-1}\left(\Delta\omega_j \vec{k} + \omega_1 \vec{j}\right), \tag{1}$$

which is described by a system of equations (for simplicity, the relaxation processes are not taken into account):

$$\dot{m}_{xj} = \Delta\omega_j m_{yj} - \omega_1 m_{zj}, \quad \dot{m}_{yj} = -\Delta\omega_j m_{xj}, \quad \dot{m}_{zj} = \omega_1 m_{xj}. \quad (2)$$

$\vec{k}, \vec{j}$ are unit vectors of the RCS along $z$ and $y$ axes, correspondingly.

From the solution of system (2) under the equilibrium initial conditions $m_{xj} = m_{yj} = 0$, $m_{zj} = m$ [2] it follows that the longitudinal nuclear magnetization after the pulse is equal to

$$m_{zj} = m\left(\sin^2\theta_j \cos\Delta\omega'_j t + \cos^2\theta_j\right), \quad (3)$$

where $m$ is the equilibrium value of the nuclear magnetization, $\theta_j$ is the angle between $\vec{H}_e$ (1) and $z$ axes and $\sin\theta_j = \omega_1/\Delta\omega'_j$.

Let us consider the effect of the second (reading) RF pulse in conditions

$$T_2^* \ll T_2 < \tau_{12} < T_1,$$

where $T_2$ is the transverse irreversible relaxation time and $T_2^*$ characterizes the transverse reversible relaxation time ($T_2^* \sim 1/\Delta^*$, where $\Delta^*$ is the half-width of half-maximum of the inhomogeneously broadened NMR line). At implementation of these conditions after elapsing time $\tau_{12}$ before the application of reading RF pulse it remains only the longitudinal component of nuclear magnetization

$$m_{zj}(\tau_{12}) = m\left(\sin^2\theta_j \cos\Delta\omega'_j \tau_1 + \cos^2\theta_j\right).$$

A sufficiently short second RF pulse, during which the relaxation processes and inhomogeneous broadening do not play role, turns the longitudinal magnetization around the $x$ axis on angle $\alpha$ so that just after the application of reading pulse the transverse component of magnetization is equal to

$$m_{yj}(\tau_{12}) = m\sin\alpha\left(\sin^2\theta_j \cos\Delta\omega'_j \tau_1 + \cos^2\theta_j\right). \quad (4)$$

The further calculation of the TPSE intensity is similar one carried out in [2]. The contribution to the TPSE intensity gives the first term reflecting the dependence of $z$-component of $j$-th isochromate magnetization on the phase of its transverse component in the time moment, of the extinction of the first pulse $\varphi'_j = \Delta\omega'_j \cdot \tau_1$. Allowing for the fact that the SPE formation mechanism is the most effective at the nonresonant excitation of spin system when the condition $|\Delta\omega_0| \gg \omega_1$ [2] is fulfilled one could obtain for the TPSE intensity the expression

$$I_{st} \approx \frac{1}{2} cm\eta \sin\alpha \left|\frac{\omega_1}{\Delta\omega_0}\right|^2 \cdot \exp\left\{-2\frac{\tau_1}{T_2}\right\} \exp\left\{-\frac{\tau_{12}}{T_1}\right\}, \quad (5)$$

where $c$ is a coefficient that takes into account the geometry of experiment, the time of TPSE appearance, if time is reckoned from the end of the auxiliary pulse, is approximately equal to the duration of the first RF pulse $\tau_1$.

Thus, optimum conditions of observation of TPSE of the reading RF pulse should be resonant and of $90^\circ$, similar to Hahn mechanism [1].

For the intensities of TPSE and SPE are valid the following simple assessments [2]: $I_{TPSE} \sim \left|\frac{\omega_1}{\Delta\omega_0}\right|^2$ and $I_{SPE} \sim \left|\frac{\omega_1}{\Delta\omega_0}\right|^3$, correspondingly.

Earlier the TPSE for the considered ease was observed for $Eu^{151}$ nuclei in a polycrystal sample of ferrite-garnet $Eu_3Fe_5O_{12}$ at temperature 1.7 K [8]. The calculation carried out in frames of nonresonant mechanism [2] showed that for a sufficiently short reading RF pulse of the TPSE it could be observed three more spin echo signals in time instants $\tau_{12}$ and $\tau_{12} + \tau_1$, formed by the traditional Hahn mechanism. The echo signals intensity in instants $\tau_{12}$ and $\tau_{12} + \tau_1$ are proportional to $\left|\frac{\omega_1}{\Delta\omega_0}\right|$, but the echo intensity in instant $\tau_{12} - \tau_1$ is proportional $\left|\frac{\omega_1}{\Delta\omega_0}\right|^2$, small and for this reason was not observed in work [8].

The TPSE signals were observed in powdered $Fe^{57}Fe$, $Ni^{61}Ni$ and $Fe^{51}V$ samples [9]. The TPSE signal at $\tau_1 \leq \tau_2$ was observed at the distance $\tau_1$ from the back front of the second pulse and its position was independent of the time interval $\tau_{12}$ between the exciting pulses. In case from the front of the second pulse at $\tau_1 > \tau_2$ two components were observed at the distance $\tau_1$ from its back front. The SPE signal in this work was not observed. In work [10] in the same systems it was studied also the signal of main or multiple echo possessing the multicomponent structure and linear dependence of the appearing time of one of its main components ($C_2$ component in designations of work [10]) on the duration of exciting pulse. It was noted unusually short relaxation time of this component as compared with other one ($C_1$ component) of multiple echo, which coincides with the signal of usual TPE in the limit of short pulses. The nature of multiple echo components was studied in work [11] on the example of $^{57}Fe$ NMR in lithium ferrite where it was simultaneously intensive signals of SPE, TPSE and multiple echoes ($E_1$, $E_2$, $C_1$ and $C_2$ signals [10]). In work [11] it was revealed the synchronous change of $E_1$, $E_2$, and $C_2$ echo signals intensities depending on $\tau_1$, pointing to the related mechanisms of their formation. Besides it, it was established the nature of short relaxation time for $C_2$ component. This time appeared to be close to the relaxation time of $E_1$ component which was, as it was earlier established, almost on two orders of value shorter as compared with transverse relaxation time $T_2$, determined from the Hahn TPE method ($T_2^S$ =60 μs and $T_2$=1200 μs, correspondingly [12]). Further, in work [13] it was studied the dynamics of $E_1$ and $C_1$ components in polycrystalline sample of ferromagnetic $Co$ where the ratio of relaxation times of this components appeared to be different and was defined by the longer relaxation time of SPE in cobalt, formed by the distortion mechanism: namely, $T_2^S = 0.4 \div 0.7\, T_2$, where $T_2 = 60$ μs.

In this work we present the results of the similar investigations of TPSE nature in polycrystal samples of lithium ferrite, cobalt, intermetal $MnSb$ and half metal $Co_2MnSi$ where signals of SPE could be formed by nonresonant or distortion mechanisms of by the both mechanisms simultaneously.

It will be studied in more details the formation mechanisms of SPE and TPSE in systems where the investigated SPE signals where formed by the distortion mechanism.

In work [14] it was studied experimentally and theoretically the peculiarities of the distortion mechanism at formation of SPE in polycrystalline films of *Co* at helium temperatures. The influence of transition processes on the fronts of RF pulse was visualized directly by the supplying signal from the spectrometer resonator to the input of the high-frequency pulsed oscilloscope S-75. But in work [13] it was suggested other, more effective method for visualization of the influence of transition processes on the RF pulse fronts by application of additional magnetic video-pulse (MVP) field, allowing more detailed study of RF pulse distortion structure. The method consists in the comparison of timing influence of MVP on SPE and TPSE signals and is calculated in Fig. 4 in work [13] on the example of lithium ferrite.

The analysis of timing diagrams shows the definite analogy between the MVP influence on the SPE and TPSE signals in lithium ferrite.

The corresponding timing diagrams for the MVP influence for *Co* and $Co_2MnSi$ are investigated in work [15] and confirm conclusions of work [13] for magnets where the SPE signals are formed by the distortion mechanism.

On the basis of obtained picture of the MVP influence one could make conclusion on the definite duration of transition process regions in the resonant circuit of spectrometer and assume that at decreasing of the PF pulse duration it is possible their overlapping resulting in the simplification of pulse structure, approaching it to rectangular shape when disappeared the edge structure of RF pulse revealing at application of MVP [15]. In this range of RF pulse duration it could take place the change of one SPE formation mechanism on other being characteristic for systems without the observable RF pulse distortion where the SPE signal is absent at resonant excitation as it is the case in lithium ferrite [6].

With this aim it was carried out corresponding measurements of SPE and TPSE intensities in the range of RF pulse duration down to short one, close to the recovery time of the spectrometer characterizing the transient loss of its sensitivity following the RF burst which in our case constitutes value of the order of ~ 1 μs [13].

## EXPERIMENTAL RESULTS AND DISCUSSIONS

A standard phase-incoherent spin-echo spectrometer was employed for measurements in frequency range 40-400 MHz at temperature 77 K. In frequency range 40-220 MHz a standard self-excitation RF oscillator has been used. The frequency of oscillator could be gradually retuned by using a number of circuits with different inductance coils and adjustable capacities. In the frequency range 200-400 MHz it was used a commercial manufactured oscillator based on the two-wire Lekher-type line including two coils with different numbers of turns. For pulse lengths ranging between 0.1 and 50 μs a maximum RF field produced of the sample was estimated to be about 3.0 Oe, while the rise and fall times of RF pulse fronts were no more than 0.15 μs. The recovery time of the spectrometer characterizing the transient loss of its sensitivity following the RF burst, was about ~ 1 μs. For investigation of lithium ferrite the resonant system of spectrometer was modified similar to described in work [12] allowing us to increase sharply its sensitivity as compared with one in work [13].

It was used the circular discs of dielectric lithium ferrite and its solid solutions with zinc $Li_{0.5}Fe_{2.5-x}Zn_xO_4$ (0≤$\chi$≤0.25) enriched by isotope $^{57}Fe$ (96.8 %) and also hexagonal polycrystalline cobalt and half metallic $Co_2MnSi$ for NMR investigations of $^{59}Co$ and $^{55}Mn$ nuclei. Half metals are interesting for applications in spintronics [16]. The properties of all above noted samples are described in details in [13].

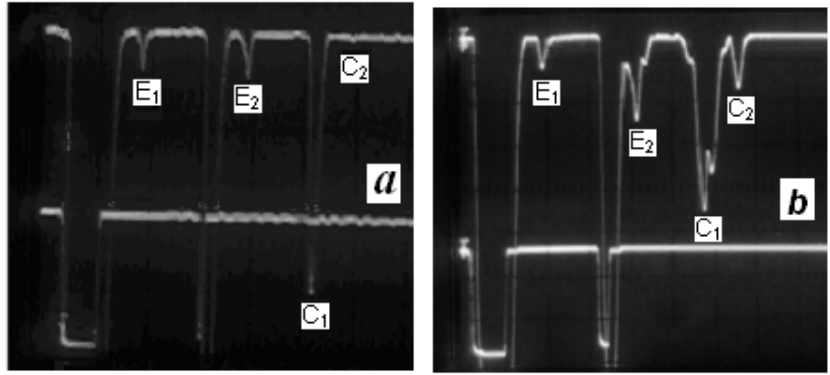

**Fig. 1.** Oscilloscope signals of SPE, TPSE and multiple echo ($E_1$, $E_2$ and $C_1$, $C_2$, correspondingly) in lithium ferrite (a) and cobalt (b). Lower beams show the RF pulse shape, amplitude and duration at:

a) $\tau_1$=14 µs, $\tau_2$=1.2 µs, $\tau_{12}$=40 µs, $f_{NMR}$=71.6 MHz, $T$=77K;

b) $\tau_1$=12 µs, $\tau_2$=2.5 µs, $\tau_{12}$=22 µs, $f_{NMR}$ = 217 MHz, $T$=77K.

In Fig.1 it is presented oscilloscope signals of SPE, TPSE and multiple echo following the applications of two pulse of optimal power in lithium ferrite and cobalt, but in Fig. 2-6 it is shown the corresponding relaxation dependences for TPSE, SPE and TPE in lithium ferrite, cobalt, $MnSb$ and $Co_2MnSi$, correspondingly, obtained by changing the duration of the first wide RF pulse and at fixed value of $\tau_{12}$. It should be noted also that SPE signals in lithium ferrite (Fig. 7) and intermetal $MnSb$ were not observed in the limit of single-pulse excitation by a single RF pulse (when pulse repetition period is long as compared with $T_1$) in the conditions of our experiment.

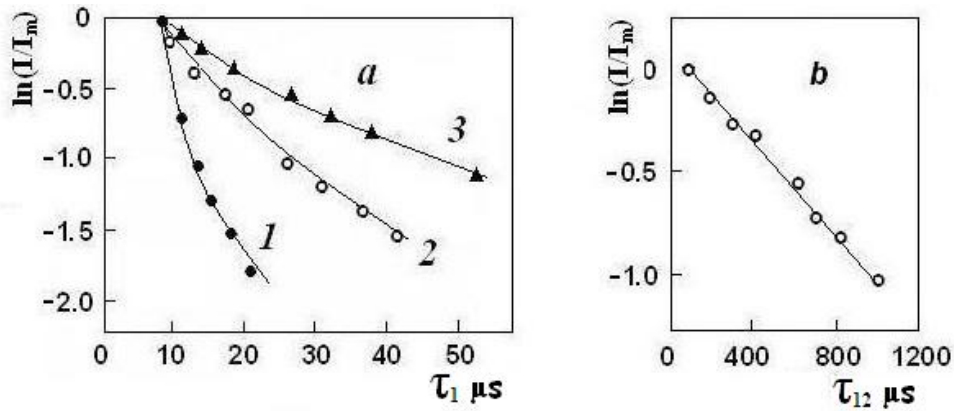

**Fig. 2**. a) Intensities of TPSE intensity at single excitation by a pair of RF pulses (1), and also the TPSE (2) and TPE (3) at the repetition rate of pair and ,correspondingly single RF pulses $f_p$ =100 Hz depending on the duration of the first RF pulse $\tau_1$ (at $\tau_2$=1 µs) in lithium ferrite. The frequency of $^{57}Fe$ NMR $f_{NMR}$ =71.7 MHz and $T$

=77 K. b). The intensity of TPE depending on the duration of time interval between RF pulses. The repetition rate of pairs of RF pulses $f_p$ =100 Hz and $\tau_1 = \tau_2 = 1\mu s$, $f_{NMR}$ =71.7 MHz and $T$ =77 K.

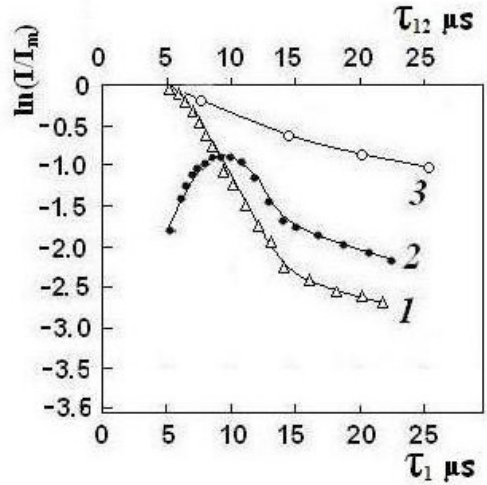

**Fig. 3**. Intensities of TPSE (1) and SPE (2) depending on the duration of the first RF pulse $\tau_1$ (at $\tau_2$=1 μs), and also TPE (3) on the interval between pulses $\tau_{12}$ (at $\tau_1 = \tau_2 = 1$ μs) in cobalt at $f_{NMR}$ =217 MHz, $T$ =77 K

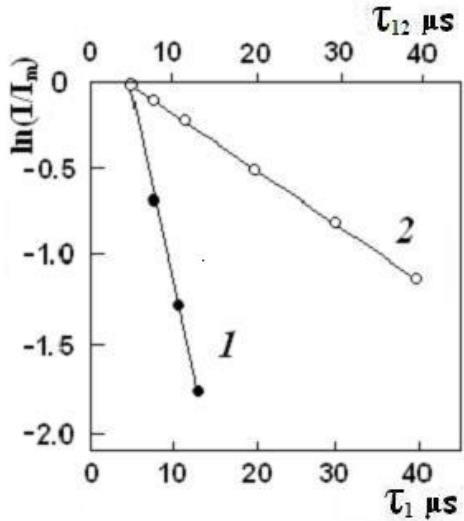

**Fig. 4**. Intensities of TPSE of $^{55}Mn$ nuclei in $MnSb$ depending on the duration of the first RF pulse $\tau_1$ (at $\tau_2$=1 μs, $\tau_{12}$=15 μs) (1) and the TPE depending on the interval between RF pulses $\tau_{12}$ (at $\tau_1 = \tau_2 = 1$ μs) (2) at $f_{NMR}$=225 MHz and $T$ =77 K.

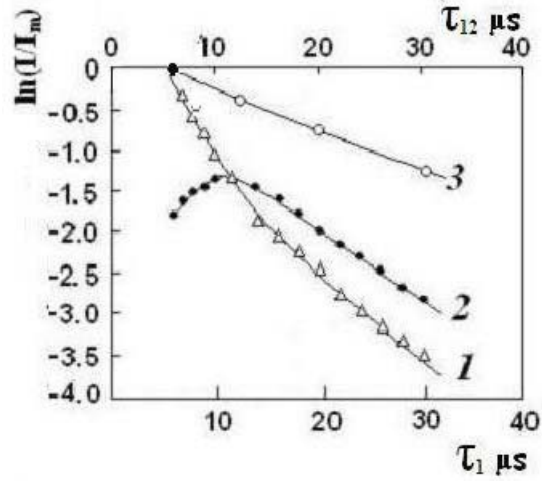

**Fig. 5**. Intensities of TPSE (1) and SPE (2) depending on the duration of the first RF pulse $\tau_1$ (at $\tau_2 =1$ μs, $\tau_{12} =27$ μs) and also TPE (3) on the interval between RF pulses $\tau_{12}$ ($\tau_1 = \tau_2 = 1.2$ μs) in $Co_2MnSi$ for $^{59}Co$ nuclei, at $f_{NMR} =142$ MHz, $T =77$ K.

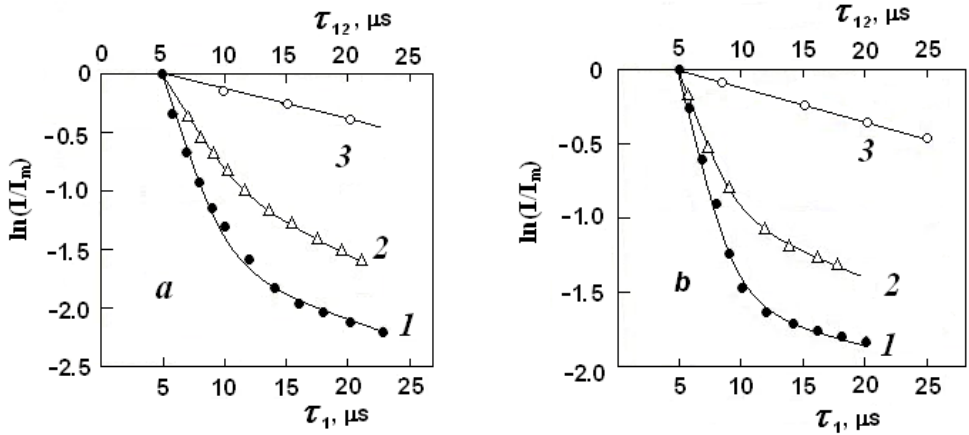

**Fig. 6**. Intensities of TPSE (1) and SPE (2) depending on the duration of the first RF pulse $\tau_1$ (at $\tau_2 =1$ μs, $\tau_{12} =30$ μs), and also TPE (3) on interval between RF pulses $\tau_{12}$ (at $\tau_1 = \tau_2 = 1.2$ μs) at repetition rate of RF pulse pair: a) 5 Hz and b) 50 Hz in $Co_2MnSi$ for $^{55}Mn$ nuclei, $f_{NMR} =353$ MHz, $T =77$ K.

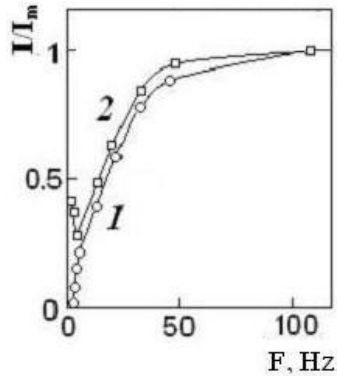

**Fig. 7**. Intensities of SPE (1) and TPSE (2) depending on the repetition rate $F$ of pulse train in lithium ferrite. For SPE - $f_{NMR} = 71$ MHz and $\tau_p = 8.5$ μs ($\tau_1 = \tau_p$ – RF pulse duration), and for TPSE - $f_{NMR} =71$ MHz, $\tau_1 = 15$ μs, $\tau_2 = 1.2$ μs and $\tau_{12} = 60$ μs [11].

Thus, the observed in the present work fast relaxation processes for TPSE are apparently. As seen from the corresponding graphs (Fig. 3, 5), the assumption on the possibility of changing the SPE formation mechanisms at decreasing the duration of first RF pulse could turn out right, as the experimental dependences of the TPSE intensity on the duration of the first RF pulse do not have any peculiarities in the range of $\tau_1$ when the SPE intensity starts to reduce in the case of spin echo of $^{59}Co$ nuclei. Besides it, the observed relaxation dependences could be approximated by the two relaxation processes: the first one being related with the distortion mechanism, and the second one characterized by the much shorter relaxation time most probably could be related with the nonresonant mechanism [2]. Let us note, that the characteristic hump-like shape of SPE signals is observed only at recording echo signals from $^{59}Co$ nuclei in $Co$ and $Co_2MnSi$, but it is absent in the case of SPE signals from $^{55}Mn$ and $^{57}Fe$ nuclei what could be caused by the contribution of anisotropic part of HF interaction which is much stronger in the case of $^{59}Co$ nuclei.

Similar ones firstly experimentally revealed in work [11] and they could be understood in the frames of the nonresonant mechanism [2, 12].

The suggested approach could turn out to be useful for the clearing out the SPE formation mechanism in systems where the distortion mechanism is effective. Besides it, this approach could help to advance in the understanding of the problem in what degree the SPE properties in such systems are defined by the physical properties of systems under study as example, are there metals of magnetic dielectrics – ferrites), or by the residual RF pulse distortions caused by transition processes in spectrometers.

In conclusion, let us note that we studied the TPSE in different magnets, such as lithium ferrite $Li_{0.5}Fe_{2.5-x}Zn_xO_4$ and intermetal $MnSb$ when the SPE signals are not observed in conditions of single-pulse excitation, and also on such ones where the SPE signals are observed in the same conditions and are formed by the distortion mechanisms – in ferrometal cobalt and in the half metal $Co_2MnSi$.

The TPSE relaxation processes in these magnets have two-component character. One of components has relaxation rate approximately on the order of value exceeding the transverse relaxation rate measured by TPE, but another one has the relaxation value close to the transverse relaxation rate of SPE in systems where the SPE signal is formed by the distortion mechanism.

Besides it, the experimental results point to the possible role of the HF anisotropy in the formation of the characteristic hump-like dependence of SPE from $^{59}Co$ nuclei in $Co$ and $Co_2MnSi$ on the duration of RF pulse.